# Theory of SEI Formation in Rechargeable Batteries:

# Capacity Fade, Accelerated Aging and Lifetime Prediction

Matthew B. Pinson[1] and Martin Z. Bazant[2,3]

Departments of Physics[1], Chemical Engineering[2], and Mathematics[3]

Massachusetts Institute of Technology, Cambridge, MA 02139

ABSTRACT

Cycle life is critically important in applications of rechargeable batteries, but lifetime prediction is mostly based on empirical trends, rather than mathematical models. In practical lithium-ion batteries, capacity fade occurs over thousands of cycles, limited by slow electrochemical processes, such as the formation of a solid-electrolyte interphase (SEI) in the negative electrode, which compete with reversible lithium intercalation. Focusing on SEI growth as the canonical degradation mechanism, we show that a simple single-particle model can accurately explain experimentally observed capacity fade in commercial cells with graphite anodes, and predict future fade based on limited accelerated aging data for short times and elevated temperatures. The theory is extended to porous electrodes, predicting that SEI growth is essentially homogeneous throughout the electrode, even at high rates. The lifetime distribution for a sample of batteries is found to be consistent with Gaussian statistics, as predicted by the single-particle model. We also extend the theory to rapidly degrading anodes, such as nanostructured silicon, which exhibit large expansion on ion intercalation. In such cases, large area changes during cycling promote SEI loss and faster SEI growth. Our simple models are able to accurately fit a variety of published experimental data for graphite and silicon anodes.

The study of rechargeable batteries, especially lithium-ion cells, is currently of huge scientific and technological interest.[1,2] In addition to experimental studies aiming to engineer superior batteries, a substantial amount of modeling has been undertaken to understand the electrochemical processes that take place during battery use.[3] Most theoretical modeling (e.g. for LiFePO$_4$ cathodes[4–10]) has focused on the performance of a battery during discharge, while much less progress has been made in the modeling of battery lifetime.  Rapid degradation is often the limiting factor in developing new electrode materials (e.g. polysulfide shuttling in sulfur cathodes[11,12] or fracture in silicon anodes[13,14]), but practical Li-ion batteries necessarily exhibit slow degradation, over hundreds or thousands of cycles.  Such slow capacity fade usually arises from irreversible electrochemical processes, which gradually compete with reversible lithium intercalation in the electrodes.

The availability of a simple, but accurate, mathematical model of capacity fade and lifetime statistics could significantly accelerate battery development and commercialization. Such a model could be used to enable lifetime prediction from short-term tests, to design accelerated aging protocols, to inform reliability assessments, and to guide efforts to reduce degradation through materials design and system optimization.  In order to reach these goals by rational design, it is essential for the model to be based on the underlying physics of degradation, rather than the fitting of empirical circuit models.

The most common and fundamental source of capacity fade in successful Li-ion batteries (which manage to resist degradation over hundreds of cycles) is the loss of lithium to the solid-electrolyte interphase (SEI), which typically forms at the negative electrode during recharging.  Initially, SEI formation protects the electrode against solvent decomposition at large negative voltage, but over time it leads to a gradual capacity fade as the SEI layer thickens.  A solid theoretical understanding of this phenomenon will assist the design of batteries, for example by enabling the quantitative interpretation of accelerated aging tests, where a battery is cycled at a high temperature to hasten the progress of capacity fade.[15]  For such purposes, a model should be as simple as it can be without neglecting important effects.

In this work, we systematically build a theory of SEI growth and capacity fade. First, we build upon simple analytical models that assert a $\sqrt{t}$ dependence of capacity fade due to the increasing thickness of the SEI layer,[16,17] by taking into account the rate of the reaction forming SEI. Next we extend the theory to porous electrodes to examine the effect of any spatial dependence of SEI formation.[18] We then use the theory to predict battery lifetime statistics and interpret accelerated aging tests at elevated temperatures. Finally, we discuss extensions to account for additional degradation mechanisms that accelerate capacity fade, such as area changes and SEI delamination resulting from volume expansion of the active particles.

SEI formation at the negative electrode is not the only possible source of capacity loss in a lithium ion battery. For example, a substantial decrease in the surface conductivity of $LiNi_{0.8}Co_{0.15}Al_{0.05}O_2$ has been found to dominate capacity fade when this material is used as the cathode.[19, 20] Since the model presented in this work treats only SEI formation, it will not be applicable when other sources of fade dominate. Nevertheless, the model should be quite generally applicable, since SEI formation is likely to be the dominant remaining source of fade in a battery optimized for long life.

**SEI formation**

In general, SEI growth results from irreversible electrochemical decomposition of the electrolyte, which competes with the desired Faradaic half-cell reaction at the electrode surface. In the case of Li-ion batteries, SEI is formed at the negative electrode because typical electrolytes are not stable at the operating potential of this electrode during charging. The product of this decomposition forms a solid layer on the surface of the active material. A huge variety of compounds has been observed in this layer,[21,22] though we largely ignore this chemical diversity.

If SEI formation were sustained throughout battery operation, it would render Li-ion batteries unusable due to the continual loss of lithium. The reason that Li-ion batteries can operate is that the SEI does not conduct electrons, and is almost impenetrable to electrolyte molecules.[21] Once an initial SEI layer has formed, the inability of electrolyte molecules to travel through the SEI to the active material surface, where they could react with lithium ions and electrons, suppresses further SEI growth.[16] Intercalation is suppressed far less, because lithium ions can easily pass through the SEI through the exchange of ions between the solvent, SEI compounds and the lithium intercalated in the active material.[21] Thus the battery is able to experience many charge-discharge cycles with little additional SEI build-up. Figure 1 provides a schematic diagram of the competing reactions.

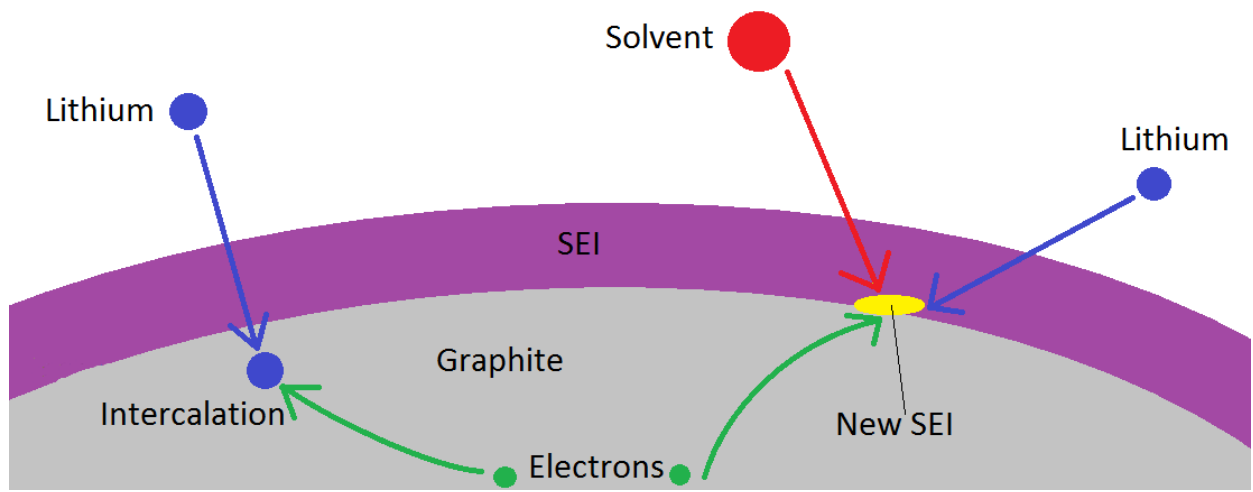

**Figure 1: The desired electrochemical reaction is the intercalation of lithium, but lithium can also react with components of the electrolyte to form a solid-electrolyte interphase.**

**Single particle model**

This work models capacity fade by considering only the loss of lithium to the SEI on the negative electrode, assuming that other sources of capacity fade can be neglected. We begin with the simplest possible model, which makes some further assumptions:

- *Uniformity.-* We assume that the amount of lithium in the SEI can be parameterized by a single variable $s$, the thickness of the SEI. This thickness is assumed to be independent of location within the negative electrode, meaning that the electrode is homogeneous and is thin enough in the direction of lithium propagation that significant concentration gradients do not form. This assumption is justified *a posteriori* below using a more sophisticated porous electrode model.

- *Negligible solid deformation.-* All intercalation compounds experience local expansion (and in some cases, also anisotropic contraction[9, 23] due to lattice mismatch) on lithium insertion. If the lattice mismatch is not too large (<5% strain), as in graphite[24, 25] then the solid expands reversibly with a concomitant, small increase in the internal surface area. The area available for SEI growth is thus fluctuating, but for practical battery electrodes it is reasonable to assume a constant surface area as we do in our initial model. The SEI layer also experiences small enough strain that it is likely to remain uniformly adhered to the surface.

For certain emerging electrode materials, it is well known that solid deformation plays a major role in limiting the performance and cycle life by accelerating the rate of SEI formation. In particular, silicon is a very attractive high-energy-density anode material for Li-ion batteries, but lithium insertion causes isotropic volume expansion by a factor of four.[26] During expansion and contraction of the active particles, the area for SEI growth changes, and the SEI layer is placed under large cyclic stresses that can lead to delamination. It is straightforward to include these effects in our single particle model, but we postpone these considerations to the end of the paper, since they are not relevant for most practical cells.

- *Constant base reaction rate.-* We assume that, if electrolyte concentration were constant at the reaction surface, the rate of SEI formation would remain constant. At first glance, this assumption may seem shaky because it ignores any dependence of SEI formation rate on the state of charge of the electrode and even on the magnitude and direction of the intercalation current. However, the SEI formation rate should depend on these factors only through its dependence on potential, and open circuit potential is close to constant over a wide range of lithium concentration in graphite.[27] The local potential will not differ too much from the open circuit potential provided the battery is not cycled at an extremely high rate, so the potential at the graphite surface remains rather constant.

- *First-order solvent decomposition kinetics. -* We make the further assumption that the SEI formation rate is proportional to the concentration of the reacting electrolyte species. The most important part of this assumption, particularly in the limit of long times, is the observation that the formation rate must go to zero with reactant concentration.

Experimental data (see below for a comparison of this model with experimental results[28,29]) suggest that the large time limit is reached within a few days.

- *Linear solvent diffusion in the SEI.-* The very slow transport of electrolyte molecules through the SEI is the rate-limiting part of the SEI formation process. We assume that this transport can be modeled as being driven by a concentration gradient between the outer and inner surfaces of the SEI, and that the concentration difference is directly proportional to the thickness of the SEI and the reaction rate consuming electrolyte at the inner surface.

Under these assumptions, the SEI thickness can be described by two equations. The first,

$$\frac{ds}{dt} = \frac{Jm}{\rho A}, \quad (1)$$

relates the rate of increase of SEI thickness to the reaction rate $J$, the mass $m$ of SEI formed by a single reaction, the density of SEI $\rho$ and the graphite surface area $A$. We arbitrarily choose lithium fluoride as a typical SEI component:[21] its molar mass is 26 g/mol and density is 2.6 g/cm$^3$. The reaction rate $J$ can be described by

$$J = kA(c - \Delta c), \quad (2)$$

where

$$\Delta c = \frac{Js}{AD}$$

is the concentration difference between the outside and the inside of the SEI, needed to transport electrolyte molecules through with flux $J/A$. In equation 2, $k$ is a reaction rate constant, a fitting parameter in our model. $D$ is the diffusivity through the SEI of the species that reacts with lithium to form SEI, while $c$ is the concentration of this species in the bulk of the electrolyte. Since we do not know exactly which species is dominant in the continuing formation of SEI during cycling, we assume that $c$ is 1 M, a typical concentration of anions. We have made several arbitrary assumptions in order to present $D$ and $k$ in useful units; their values should scale appropriately (e.g. with temperature) but will only be a vague guess at the correct values.

Equations 1 and 2 can be solved analytically to give the SEI thickness as a function of time:

$$s = \frac{\sqrt{2c\rho mk^2 Dt + D^2\rho^2} - D\rho}{\rho k}. \quad (3)$$

In the limit of large $t$, this becomes

$$s = \sqrt{\frac{2cmDt}{\rho}} - \frac{D}{k}, \quad (4)$$

which reproduces the $\sqrt{t}$ dependence of Ploehn et al.[16] and Smith et al.[17]

Two questions must be answered to assess the ability of this simple model to explain SEI formation: can it match experimental results, and if so, does it hide variation that may be important in some cases? We answer the first by attempting to explain experimental results using equation 4, treating $k$ and $D$ as adjustable parameters. For example, we apply the model to the results of Liu et al.[28] Inferring the SEI thickness from the percentage capacity loss (assuming an average particle radius of approximately 5 µm), we find that the approximation of equation 4 is valid even after the first charge-discharge cycle. Using this equation we find $D = 2 \times 10^{-17}$ cm²/s at 15°C and $D = 3 \times 10^{-16}$ cm²/s at 60°C. Figure 2 compares experimental data with the single particle model, as well as with the porous electrode model described in the next section. In the cell cycled at 60 °C, capacity loss between 707 and 757 cycles is much larger than is expected under the model. This demonstrates that some other loss mechanism is important in this instance. The most likely mechanism is the delamination of graphite from the current collector: figure 10 of Liu et al.[28] shows delamination for this cell.

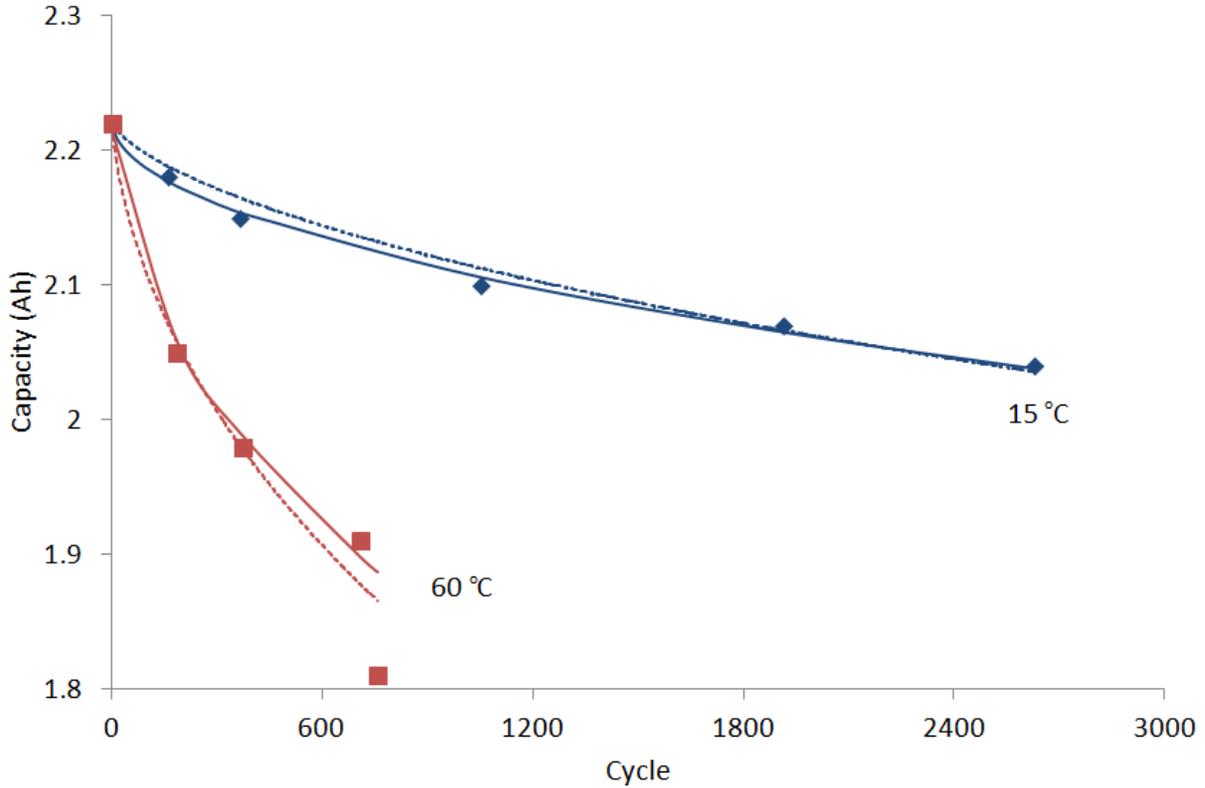

Figure 2: The single particle model (solid line) and porous electrode model (dotted line) both closely fit experimental data (points). Data from Liu et al. (28) with a lithium iron phosphate opposite electrode.

The second question, of whether the simple model obscures a variation in a way that cannot be deduced directly by comparison with experimental results, is equally interesting. Here we focus on the possibility that SEI forms non-uniformly through the electrode, due to the spatial variation in the concentration of various species in the electrode during charging. We examine this possibility using a porous electrode model.

**A porous electrode model**

We now present a one-dimensional porous electrode model[3, 4, 30–32] that accounts for concentration gradients in the primary direction of lithium propagation, which were neglected in the single particle model described above. To our knowledge, this is the first attempt to theoretically predict the spatio-temporal distribution of SEI formation in a porous electrode, using a model in which the SEI formation rate depends on the concentrations of both lithium and the reacting species from the electrolyte. Previous work by Ramadass et al.[18] modeled SEI formation throughout the negative electrode under the assumption that the formation rate is independent of lithium ion concentration and current SEI thickness. Other work, such as that of Ploehn *et al.*,[16] has accounted for the decrease in the rate of SEI formation caused by the need for the reactants to diffuse through the already-existing layer, but ignored any spatial variation of SEI formation throughout the negative electrode.

The porous electrode model is based on a diffusion-reaction equation,

$$\frac{\partial c_i}{\partial t} = -\frac{\partial N_i}{\partial x} - \sum J(i), \quad (5)$$

which expresses mass balance in a volume-averaged sense. Here $c_i$ is the concentration of species $i$ in the electrolyte, $N_i$ is the flux of species $i$ and $\sum J(i)$ is the sum of all interfacial reactions consuming species $i$. We let $i$ range over positive and negative ions, assuming that the consumption of electrolyte molecules by reactions forming SEI is slow enough that it does not significantly affect electrolyte concentration. The fluxes $N_i$ are calculated using the Nernst-Planck equation. The rate of the reaction $Li^+_{(dissolved)} + e^- \rightarrow Li_{(intercalated)}$ is modeled using a symmetric Butler-Volmer equation:

$$J = 2i_0 \sinh\left(\frac{-e\eta}{2k_B T}\right). \quad (6)$$

Here $\eta$ is the overpotential driving the reaction: the potential difference across the surface of the active material, where the reactions take place, minus the equilibrium potential difference, i.e.

$$\eta = \varphi_{electrode} - \varphi_{electrolyte} - \Delta\varphi_{equilibrium} + JR. \quad (7)$$

The equilibrium potential for each electrode, relative to lithium, is drawn from figure 5 of Liu *et al.*:[28] these potentials are approximated by the functions

$$\Delta\varphi_{equilibrium}^{graphite} = (0.132c^{-0.425} - 0.0903)V,$$

$$\Delta\varphi_{equilibrium}^{LiFePO_4} = (-0.0039c + 3.405 - 908.954(c - 0.470)^{11})V,$$

where $c$ is the concentration of intercalated lithium, divided by the concentration of lithium in $LiC_6$. The slight dependence of $\Delta\varphi_{equilibrium}$ on $c_+$, the concentration of lithium ions, is ignored. We assume that the concentration $c$ is uniform throughout a given active material particle, that is, we neglect transport and phase transformations[33] in the solid. $R$, in the case of the graphite electrode, is the resistance of the SEI layer. This resistance is proportional to the thickness of the SEI layer: we can write this as

$$R = \rho c_{SEI}.$$

$c_{SEI}$ represents the amount of SEI present at a location in the electrode. Since the SEI volume is not fixed, this is expressed as Li atoms bound in the SEI per volume of electrode: it is not a true concentration on the smallest scale. The positive electrode is assumed to have no interfacial resistance. The exchange current density $i_0$ depends on concentration: we assume that

$$i_0 = k\sqrt{c_+} \quad (8)$$

for the intercalation, where $k$ is a constant: this is the simplest implementation of a symmetric Butler-Volmer equation. $c_+$ varies through the electrode but is assumed to be constant through the depth of the SEI layer. Capacity fade is modeled through a competing current that leads to SEI formation, with rate

$$J_{SEI} = 2i_{0SEI} \sinh\left(\frac{-e\eta_{SEI}}{2k_B T}\right):$$

this models the reaction $Li^+_{(dissolved)} + E + e^- \rightarrow LiE$, where E is some species from the electrolyte that can react with lithium, and LiE is the product, which forms the SEI layer. Overpotential is calculated in the same way as for the intercalation current,

$$\eta_{SEI} = \varphi_{electrode} - \varphi_{electrolyte} - \Delta\varphi_{SEI} + JR.$$

Here $\Delta\varphi_{SEI}$ is the equilibrium potential for SEI formation, approximately 0.8 V.[21,34,35] In this case the reaction requires not only a lithium ion but also an electrolyte molecule, so

$$i_{0SEI} = k_{SEI}\sqrt{c_+ c_s}.$$

The electrolyte concentration $c_s$ should be measured at the bottom of the SEI layer. Assuming that the concentration outside the SEI layer is constant, and also assuming linear diffusion, we can write

$$c_s = (1 - Zc_{SEI}J_{SEI})c_0,$$

where $Z$ is a constant inversely proportional to the diffusivity of electrolyte in SEI,

$$J_{SEI} = \frac{\partial c_{SEI}}{\partial t},$$

and $c_0$ is the concentration of the relevant electrolyte molecule outside the SEI; its value can be absorbed into $k_{SEI}$.

Using these equations along with standard modeling of transport,[4–6] SEI formation was modeled over many cycles. Figure 2 compares experimental[28] capacity fade with the predictions of the single particle and porous electrode models. Using the values of $\rho$, $k$, $k_{SEI}$ and $Z$ needed to fit these experimental data, capacity fade was simulated over a wide range of discharge rates. The results are plotted as a function of cycle number and of time in figure 3. It is clear that time, rather than cycling, is the dominant factor in SEI growth, under this model. This provides some justification for the single particle model, which treats SEI growth over time without considering the discharge rate.

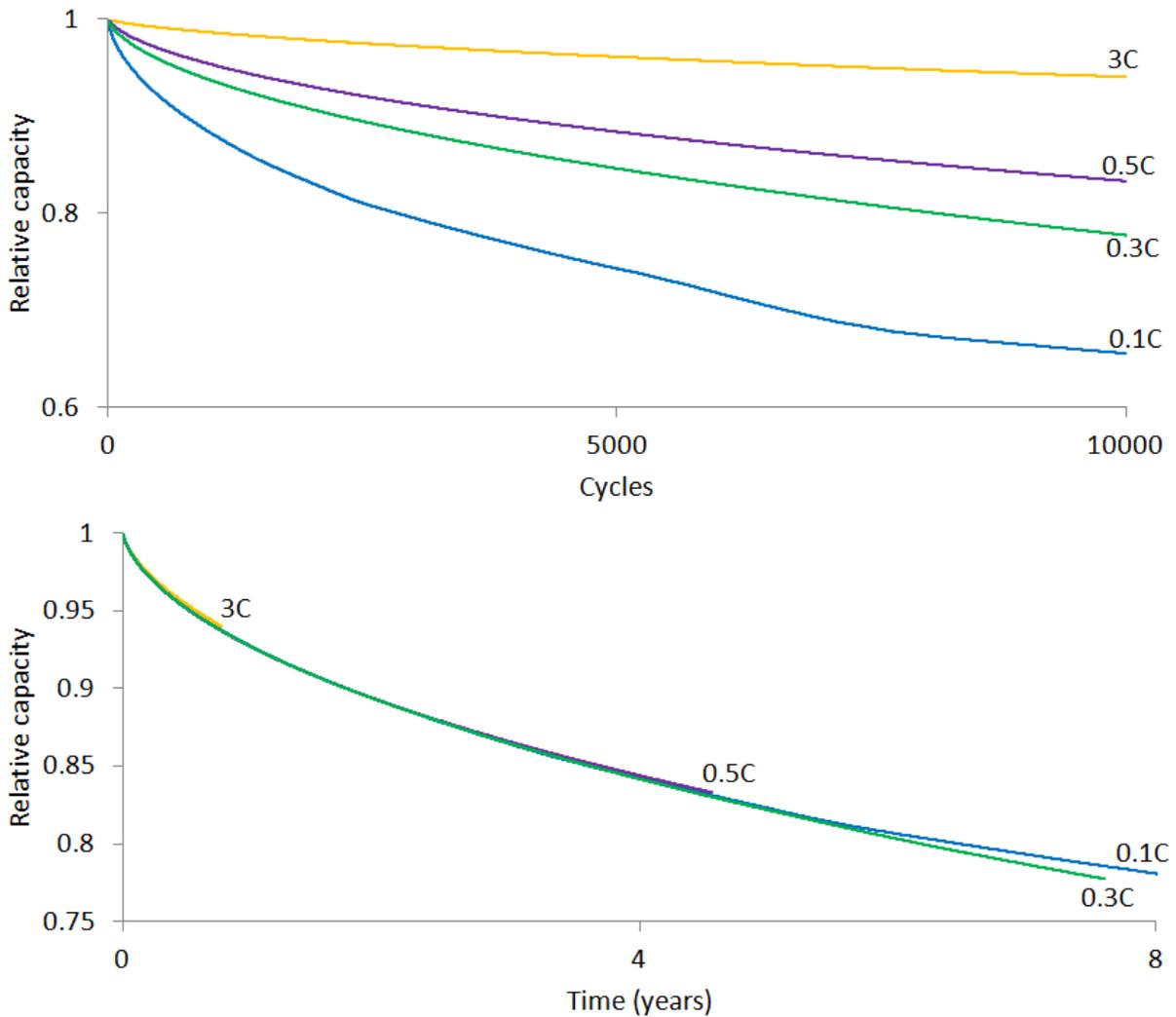

Figure 3: Simulated capacity fade clearly depends on time, not on number of cycles.

The best fit to the experimental results of Liu *et al.*[28] includes zero SEI resistivity. To ensure that this parameter choice, which simplifies computation substantially, does not interfere with the accuracy of the results, simulations were performed with a non-zero resistance. Results are shown in figure 4. The SEI resistance lowers the output voltage, but has little effect on capacity fade.

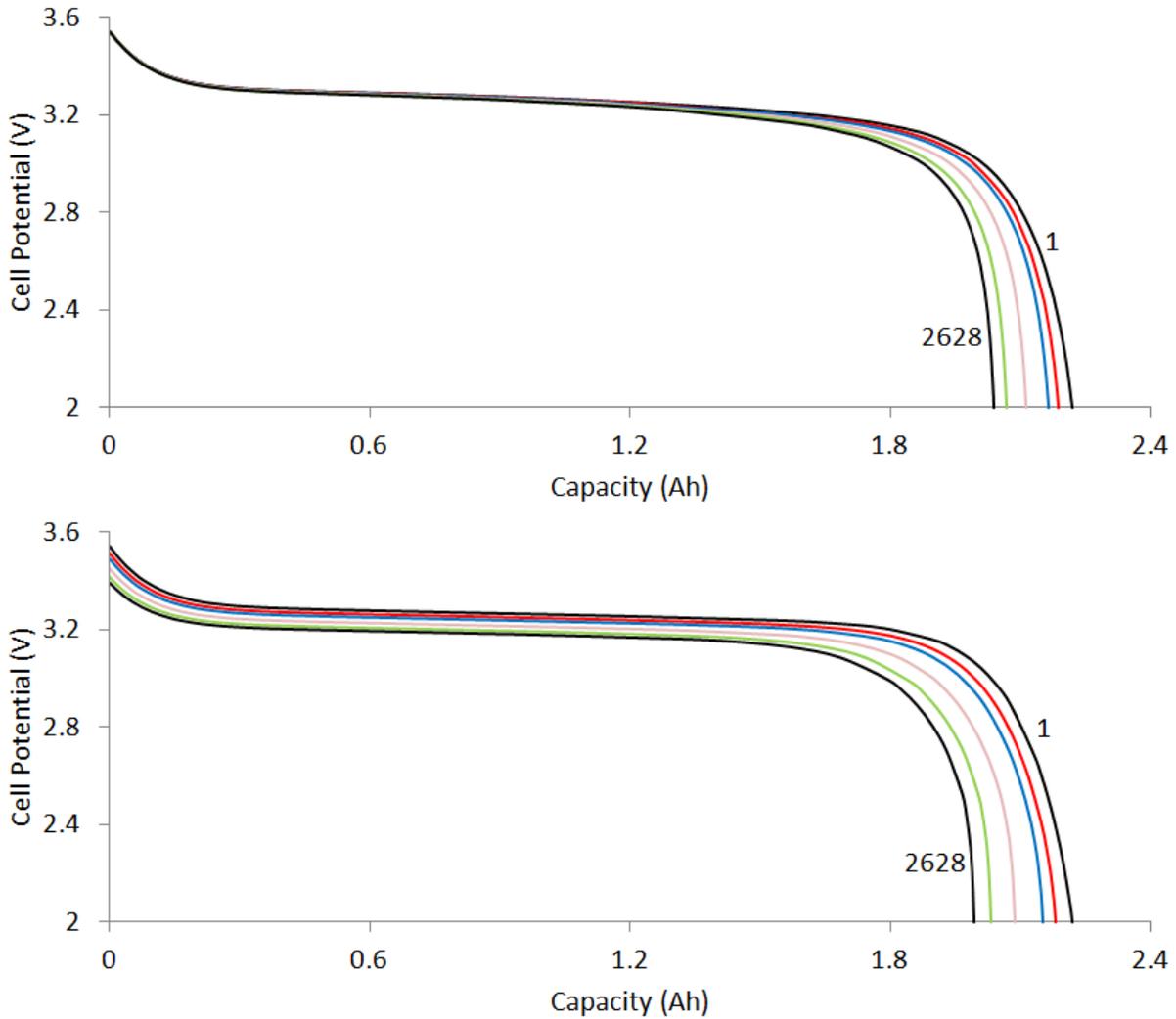

**Figure 4:** The top graph shows simulated voltage-capacity curves for the 1st, 161st, 366th, 1054th, 1914th and 2628th cycles, assuming that the SEI has infinite lithium ion conductivity. The bottom graph shows the same curves for finite lithium ion conductivity. Voltage is decreased in the latter case, while the capacity fade is not substantially affected.

An important advantage of the porous electrode model over the single particle model is that it can assess the degree of spatial variation in SEI formation. Results of the model demonstrate that in realistic battery operating conditions, SEI formation is very uniform. Only in extreme charging conditions, where the lithium concentration gradient in the electrolyte is large enough to cause substantial underutilization of the battery, do significant non-uniformities in SEI formation occur. Figure 5 compares SEI formation in an electrode of width 50 μm, which is uniform even for fast charging, with that in a 250 μm electrode, where fast charging causes a moderate degree of non-uniformity. This non-uniformity is dependent on strong depletion of the electrolyte during charging. Thus the single particle model, which ignores spatial non-uniformity, is valid in normal circumstances, but the variation of SEI through the electrode should be considered when a battery is operated in extreme conditions such as at ultra-high rate or with very thick electrodes.

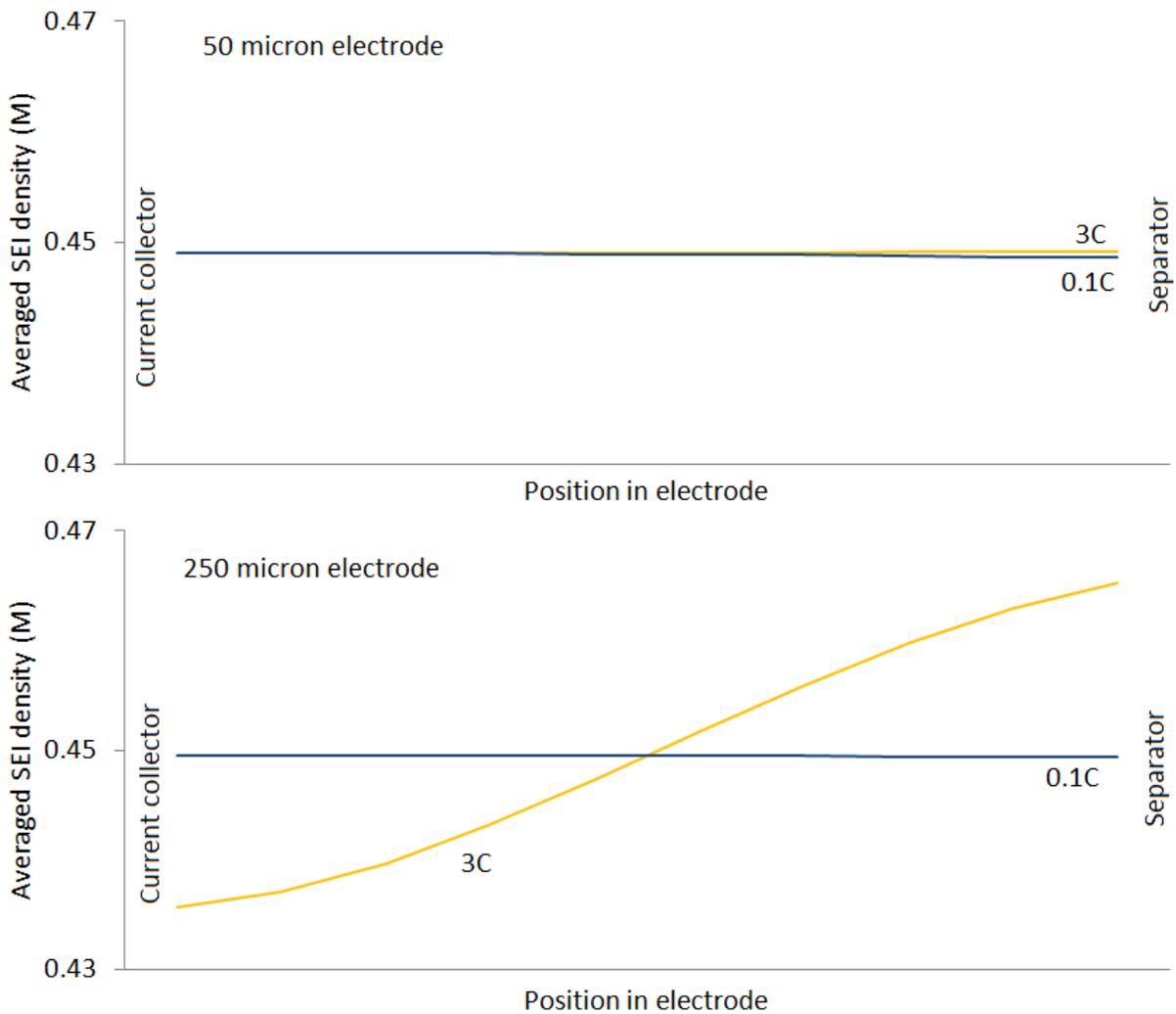

Figure 5: Computational results show that a thin electrode (top) displays no significant spatial variation in SEI formation, while a thicker electrode (bottom) does display some variation when charged at a high rate. The numbers of cycles were chosen to achieve equal average SEI density.

The porous electrode model does not explicitly account for temperature, while the single particle model assumes that the battery remains at a constant temperature. Very fast charging is likely to lead to a non-uniform temperature increase. Since SEI formation is strongly temperature-dependent, this temperature change is likely to have a larger influence on the spatial uniformity of SEI than any of the factors included in the porous electrode model.

**Temperature dependence of the diffusivity**

Since the single particle model, whose important parameter is the diffusivity of electrolyte molecules through the SEI, appears to completely explain the loss of capacity due to SEI formation, it is interesting to provide a theoretical model attempting to explain the temperature dependence of this diffusivity. The model of this work applied to the experimental data of Liu *et al.*[28] shows a clear temperature dependence of diffusivity. To quantify this dependence, we compared the single particle model with the experimental data of Smith *et al.*,[29] which span many days at temperatures from 15ºC to 60ºC. The diffusivities calculated from these data were consistent with an Arrhenius dependence,

$$D = A e^{-\frac{E}{k_B T}}, \qquad (9)$$

with activation energy $E = 0.52$ eV. Figure 6 shows the prediction of the single particle model, with diffusivities (and reaction rate constants) fit to the Arrhenius description.

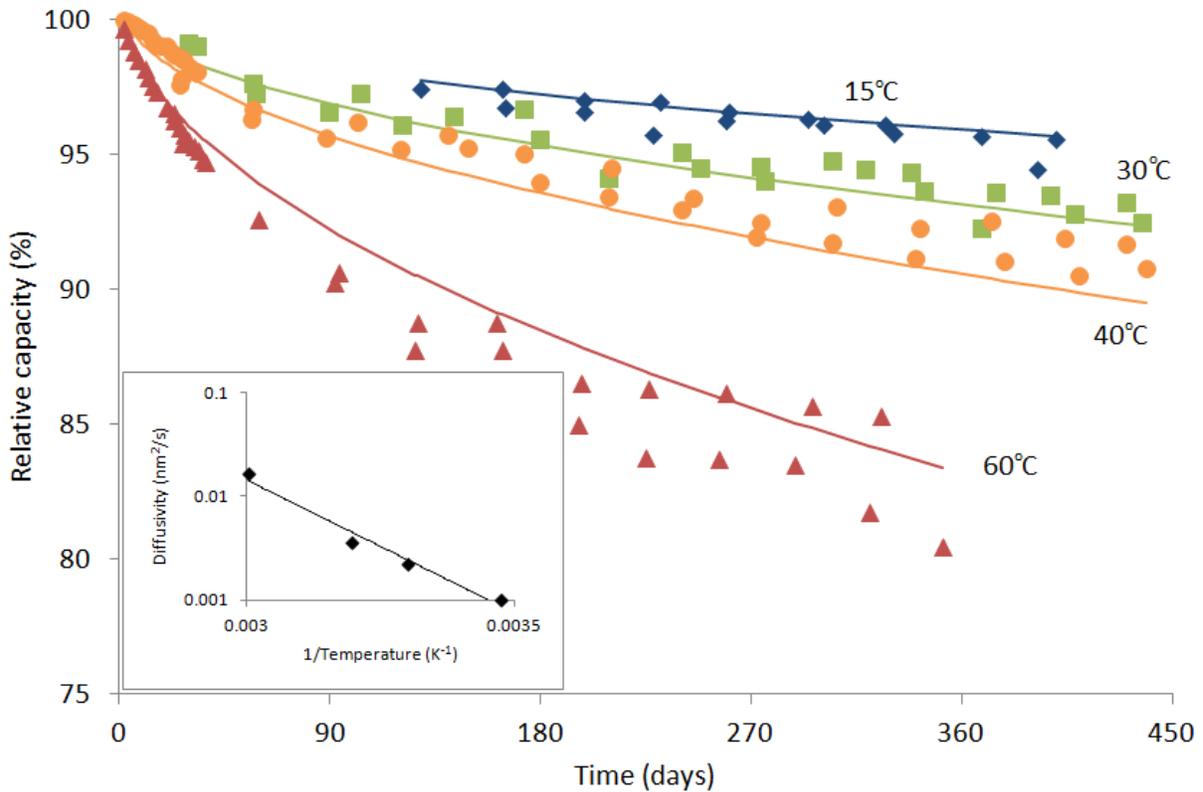

Figure 6: The single particle model accurately describes capacity fade over a range of temperatures, with diffusivity fit using an Arrhenius dependence (shown in the inset). Data from Smith et al. (29) with a lithium opposite electrode.

**Accelerated Aging**

The observed temperature dependence enables the single particle model to be used to draw quantitative conclusions from accelerated aging experiments, where an increased temperature is used to hasten capacity fade. A simple application is to estimate battery life at room temperature from data taken at elevated temperatures. The most accurate way to do this is to measure the progress of capacity fade at different temperatures, and use the results to calculate $D(T)$. As a test of this procedure, data from only the first 105 days at 30-60°C in the work of Smith *et al.*[29] were used to calculate $D(T)$ and $k(T)$, using equation 4. Results for these times and temperatures were fit to an Arrhenius equation. The resulting values of $D$ and $k$ at 15°C were used to predict capacity fade for 400 days: results are shown in figure 7. Though the agreement is not perfect, it is reasonable given that no data at the temperature or time of the predicted region were used in producing the prediction. In particular, simply assuming $s = \alpha\sqrt{t}$[16,17] (i.e.

neglecting the second term in equation 4) and using an Arrhenius equation to find $\alpha(T)$ is not as predictive: the best prediction using this method is shown by the dotted line in figure 7.

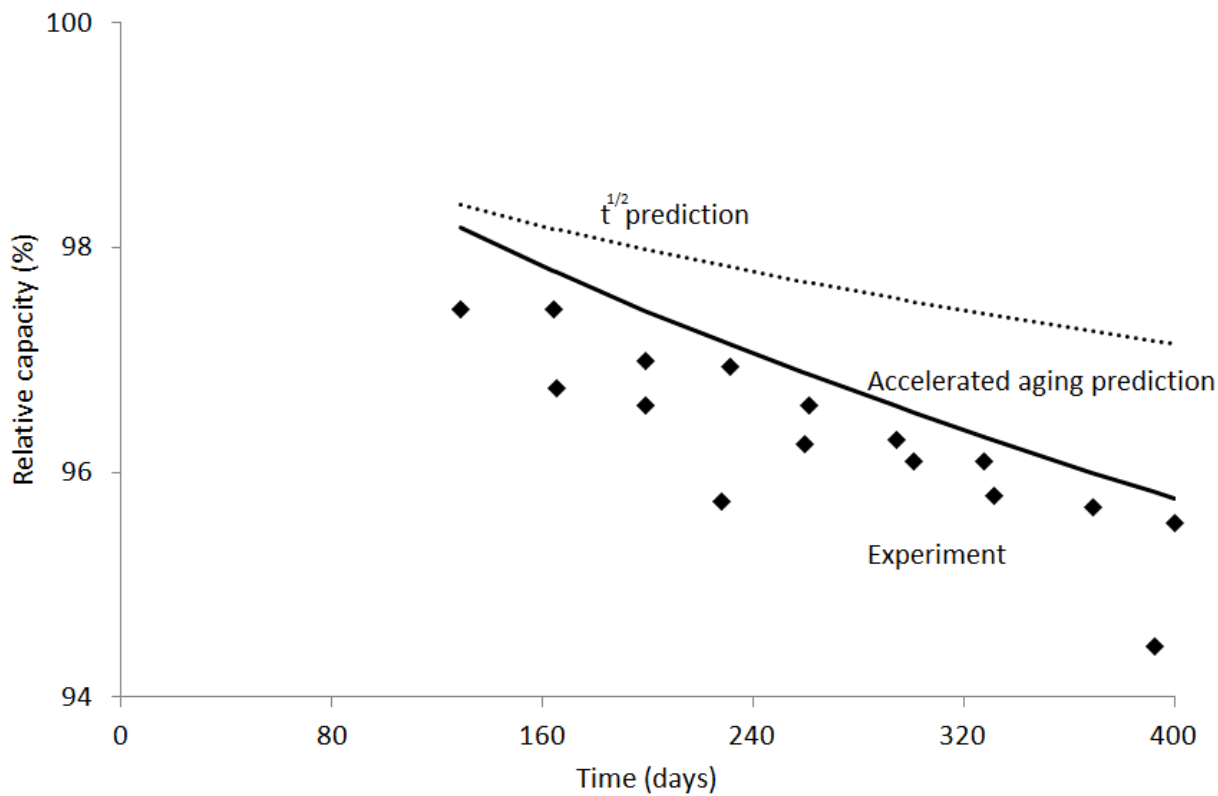

**Figure 7: The single particle model enables data from up to 105 days at 30-60ºC to be used to provide a reasonable prediction of capacity fade up to 400 days at 15ºC; assuming a simple √t dependence of capacity fade is less predictive. Data from Smith et al. (29) with a lithium opposite electrode.**

In addition, we applied the model to the data of Broussely et al.,[36] predicting capacity fade at 15ºC for 14.5 months using data at 30-60ºC for 6 months. In this case, the second term in equation 4 was statistically indistinguishable from zero, meaning that a simple $t^{1/2}$ capacity fade model would give the same result. It is possible that the second term cannot be distinguished in this case because capacity loss was not measured directly, as in Smith et al.,[29] but inferred from measured capacity. Figure 8 compares the prediction with experimental data.

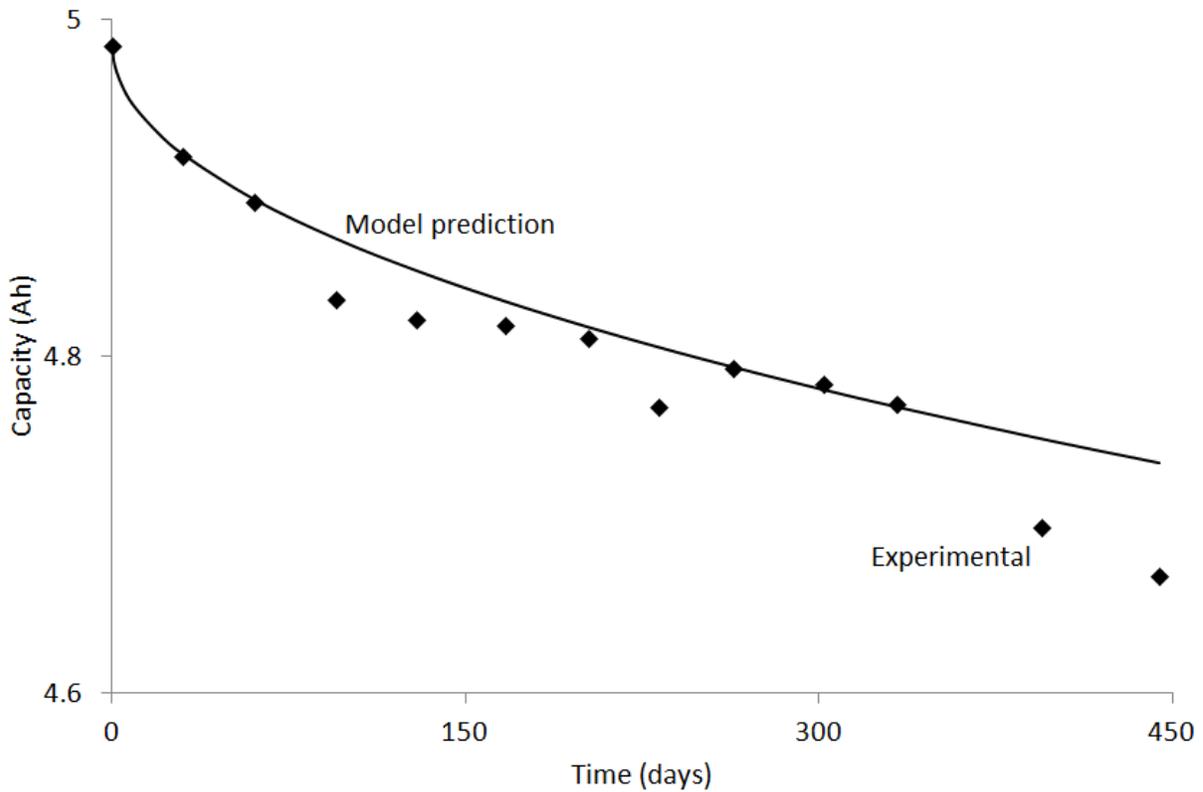

Figure 8: The single particle model, using accelerated aging data from 6 months at 30-60ºC, can be used to predict aging over 14.5 months at 15ºC. Data from Broussely et al. (36) with a lithium cobalt oxide opposite electrode.

A more sophisticated application of the single particle model would be to predict the total capacity fade of a battery subject to varying temperature. This information could be used to construct a protocol to optimize lifetime given a particular set of unavoidable temperature constraints. For example, is it possible to increase battery life by cycling it a small number of times at particularly high or low temperature, in order to produce an initial SEI with low electrolyte diffusivity? Answering this question will require experimental work to check whether the measured electrolyte diffusivity depends on the temperature history of the battery as well as the current temperature.

**Multiple reacting species**

The above analysis has considered only one species from the electrolyte reacting to form SEI. Experimental work[22] has shown that many reactions can take place to form SEI, involving a wide range of solvent species, anions or impurities. A more complete model would consider multiple reacting species, each with a characteristic $k$ and $D$. However, since experimental results very quickly reach the large time limit, we can model the process effectively simply by assuming that the $D$ implied by experiment is given by $D = \sum_i D_i$ where $i$ references the different species. The capacity fade will be dominated by the species with higher diffusivity through the SEI. This suggests that capacity fade can be reduced by constructing the cell in a way that avoids the presence of substances with high diffusivity in SEI, for instance small molecules.

The presence of multiple reacting species also provides a mechanism by which denser SEI may form at higher temperature. At higher temperature, all diffusivities are higher, and the dominance of the most easily diffusing species will be lower. If the less diffusive species form a denser SEI, as seems likely from experimental observation of denser SEI towards the bottom of the layer,[21] this will result in lower diffusivity for the remainder of the aging process and hence slower continued SEI formation.

**Lifetime prediction and statistics**

Another application of the single particle model is to understanding the statistics of battery lifetime. If the typical diffusivity of electrolyte through the SEI is $D_m$ and the battery is defined to fail when the SEI thickness reaches $s_0$, the average battery life will be $\tau_m = \rho s_0^2 / 2 c m D_m$ (we have neglected the second term in equation 4). Since a battery is made up of many, presumably identical, active particles i and the total SEI formation is proportional to $\sum_i \sqrt{D_i t}$, we expect the measured $\sqrt{D}$ to have a normal distribution, assuming the distribution of the true diffusivity $D_i$ meets the conditions of the central limit theorem. Thus the inverse of the lifetimes of a collection of batteries should be normally distributed; SEI formation as modeled here will not cause anomalously short battery lifetime.

This prediction can be compared with other predicted battery lifetime distributions, such as the Weibull distribution from the theory of extreme order statistics.[37,38] A Weibull distribution is expected when failure occurs in a 'weakest link' situation, where failure of a single element causes failure of the entire cell. Mathematically, it is the limiting distribution for the smallest outcome of a large set of independent identically distributed random variables which are bounded below, with a power law tail.[39] Additionally, a Weibull tail at short lifetime occurs in quasibrittle fracture when the system can be modeled as a bundle of fibers arranged both in series and in parallel.[40]

In order to examine the experimental lifetime distribution, we converted the data in table 2 of Park *et al.*[37] to lifetime data by defining the lifetime to be the number of cycles taken to reach a relative capacity of 85.6%, the average capacity after 400 cycles. This calculation assumed linear capacity fade between 200 and 400 cycles and slightly beyond: this is not the prediction of the model of the present work, but was used as an unbiased, simple choice that should be sufficiently accurate. Figure 9 shows a Weibull plot of calculated lifetimes. A Weibull distribution would be a straight line on this plot: this is not consistent with the data. A normal distribution of lifetimes, shown as a solid line in figure 9, is consistent with the data. If the system can be thought of as quasibrittle, we expect a Gaussian core with a Weibull tail at short lifetime.[40] The data are insufficient to establish whether this tail is present: a possible Weibull tail is represented by the dashed line in figure 9.

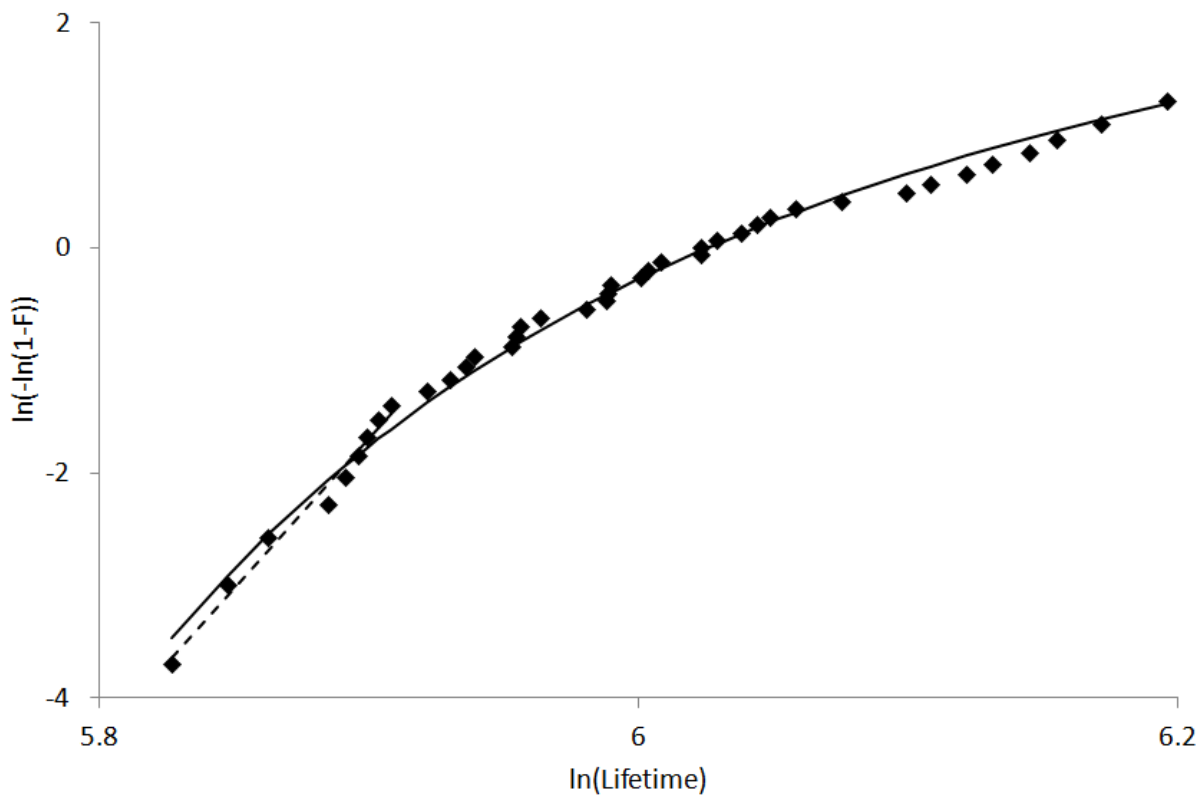

Figure 9: Lifetimes (calculated from data in Park et al. (37)) are consistent with a normal distribution (solid line), but not with a Weibull distribution, which would appear as a straight line on this Weibull plot. However, it is not possible to establish the presence or absence of a Weibull tail at low lifetime (dashed line).

**Extensions for Rapid Capacity Fade**

Until now, we have focused on gradual capacity fade due to the formation of a stable SEI film, well adhered to an electrode internal surface of nearly constant area. This is the situation in practical Li-ion cells, but various emerging technologies, which offer certain benefits, such as high energy density, suffer from much faster capacity fade related to accelerated SEI formation. The canonical and most intensely studied example is silicon, which is a very attractive anode material for Li-ion batteries due to its theoretical high energy density of 4200 mAh/g,[41] but, as a result of its enormous volume expansion upon lithiation (> 300%), degrades very rapidly. One mechanism is the loss of ohmic contacts due to mechanical deformation and fracture, but the dominant mechanism is the rapid, unstable growth of SEI. Many attempts have been made to engineer silicon nanostructures capable of accommodating the large volume expansion and stabilizing the SEI,[41–43] but no model has been developed to describe the capacity fade quantitatively, which would be necessary for the development and testing of practical cells. In this section, we extend our model to account for surface area changes and SEI delamination and apply it to published data for nanostructured silicon electrodes.

*Fresh surface.-* The single particle model assumes that the SEI is made up of a single layer growing smoothly from the surface of the active material. In a material that experiences a large volume change on lithiation, such as silicon,[26] this assumption is no longer valid. Instead, the expansion during lithiation produces fresh surface area, on which SEI can form without hindrance from an already-existing layer. As the material shrinks during delithiation, the SEI formed on the newly exposed surface will be forced into the existing layer or possibly detached from the active material. In either case, it is likely that a substantial fraction of the new area formed on the next cycle is once again freshly exposed. This leads to capacity fade at a much faster rate than would be expected from the model in which the SEI layer remains uniform and area changes are ignored.

The effect of the fresh surface area can be modeled simply by applying the result of the single particle model, equation 3 or 4, to a single charge-discharge cycle, and finding the total capacity fade by adding that of each cycle. The total SEI thickness is

$$s = \frac{t}{\tau}\left(\sqrt{\frac{2cmD\tau}{\rho}} - \frac{D}{k}\right).$$

where $\tau$ is the time spent during one cycle at a voltage where SEI forms: we assume that this is the total charge and discharge time. This simple model predicts that capacity fade is linear in time, which is precisely what is seen for full cells using silicon electrodes.[41, 44–47] Furthermore, since $t/\tau$ is the number of cycles, the rate of increase of *s* per cycle should be linear in $\sqrt{\tau}$. Figure 10 shows experimental data from figure 4c of Ji *et al.*,[44] which exhibit capacity fade of 4 mAh/g/cycle at C/15 and 2 mAh/g/cycle at C/4, consistent with this prediction. Assuming a characteristic particle size of 1 µm, this implies $D = 4 \times 10^{-15}$ cm²/s. This is two orders of magnitude higher than that found in the case of graphite, for which we offer two possible explanations. The first is that, even in the case of graphite, the new SEI formed each cycle may form primarily on the newly exposed area, though in this case the shape change is small enough that the SEI formed in the previous cycle is not fully lost. In this case, the smaller formation area means that the actual SEI thickness is higher than that calculated assuming the SEI forms evenly over the entire active surface, since capacity loss, which is measured, is proportional to SEI volume. This would mean that the true value of $D$ is higher than that calculated under the assumptions of the simple model. The second possibility is that the expansion of silicon is large enough to disturb the structure of the SEI, even within a single cycle. Such a disturbance might increase SEI porosity and is likely to increase $D$.

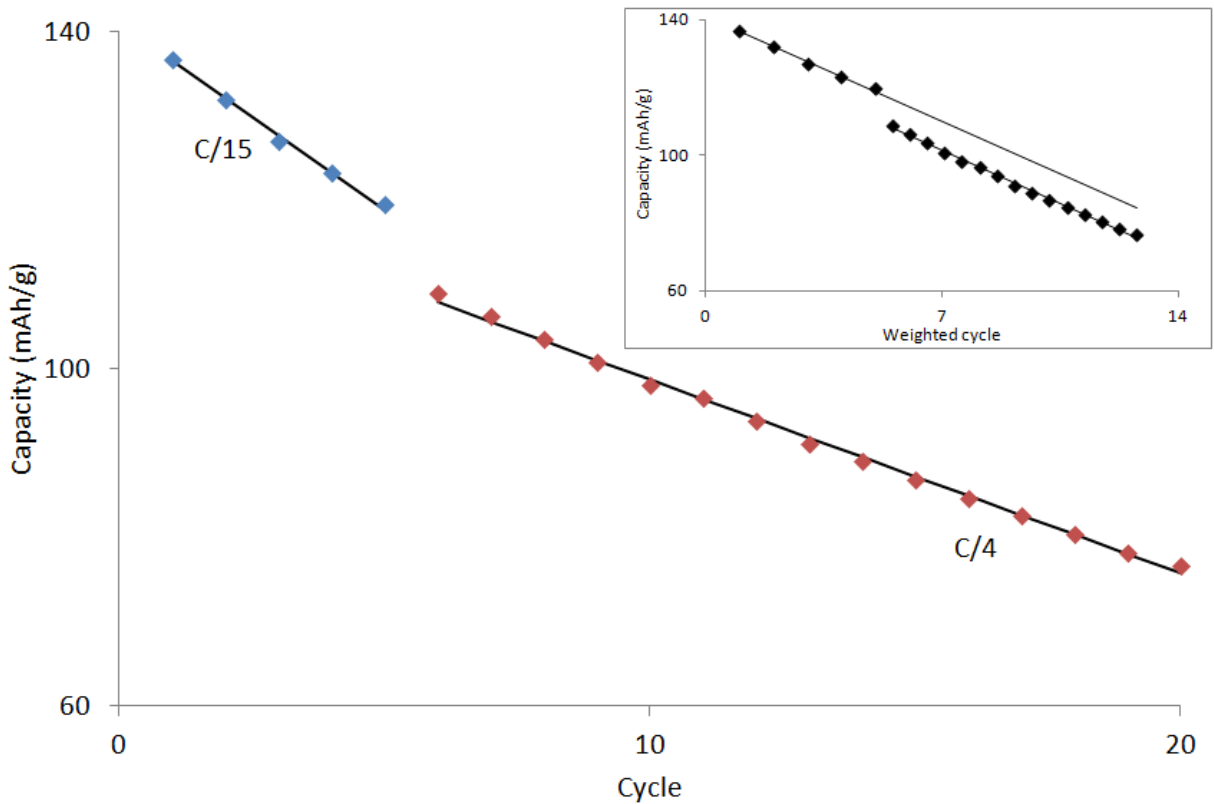

Figure 10: A full cell with a silicon anode shows linear capacity fade at two charge-discharge rates. The main graph plots capacity against cycle number. The inset plots capacity against cycle number weighted by the square root of the charge-discharge time, as the model of this work predicts that capacity fade is proportional to this quantity. The experimental data, from Ji et al., (44) match the theory very well, after accounting for the constant offset due to the chance in charge-discharge rate, which is well understood though not explicitly modeled in this work.

It is vital that full cell, not half cell, aging data are used to assess the impact of SEI growth on capacity fade. SEI growth leads to capacity fade due to the consumption of lithium that would otherwise be available for cycling. If an electrode is cycled against a lithium foil electrode containing excess lithium, this loss will not be observed, potentially leading to an overestimate of battery life. Indeed, experimental results demonstrate a substantially shorter lifetime for a full cell than a half cell using a silicon electrode: figure 4 of Baranchugov et al.[46] shows this particularly clearly. Considering only half cell data could lead to a falsely optimistic assessment of the lifespan of a silicon anode.

*Unstable SEI.-* Another situation in which SEI formation could progress faster than suggested by the single particle model is when SEI is lost from the layer, by delamination or dissolution in a form from which Li remains unavailable. Significant SEI removal is known to occur in silicon anodes, as confirmed by direct observation.[42] Assuming that this loss is not so great as to reduce the layer thickness to zero or almost zero, it can be modeled by including a loss term in the differential equation for SEI formation:

$$\frac{ds}{dt} = \frac{Jm}{\rho A} - f(s), \qquad (10)$$

where $f$ is a general function characterizing the dependence of the SEI loss rate on SEI thickness. The rate of capacity loss remains

$$\frac{dQ}{dt} = -J, \qquad (11)$$

and the dependence of $J$ on $s$ is unchanged. This alters the long time behavior of capacity. Assuming that $f$ is an increasing function of $s$ (strictly all that is necessary is that the right hand side of equation 10 is ultimately negative; in particular, a constant $f$ is sufficient), $s$ will eventually reach some maximum $s_m$. Once this maximum SEI thickness is reached, the rate of capacity loss will be $J(s_m)$. The overall result is a capacity fade rate that is linear due to reaction limitation for a short time, before moving through a $t^{1/2}$ transition region to a second linear region with a lower rate. This behavior, which could be expected for a system where the charge-discharge process places a greater burden on the SEI layer than that in graphite but less than that in silicon, is represented in figure 11. In such a situation, capacity fade could be reduced not only by decreasing $J(s)$, but also by decreasing $f$.

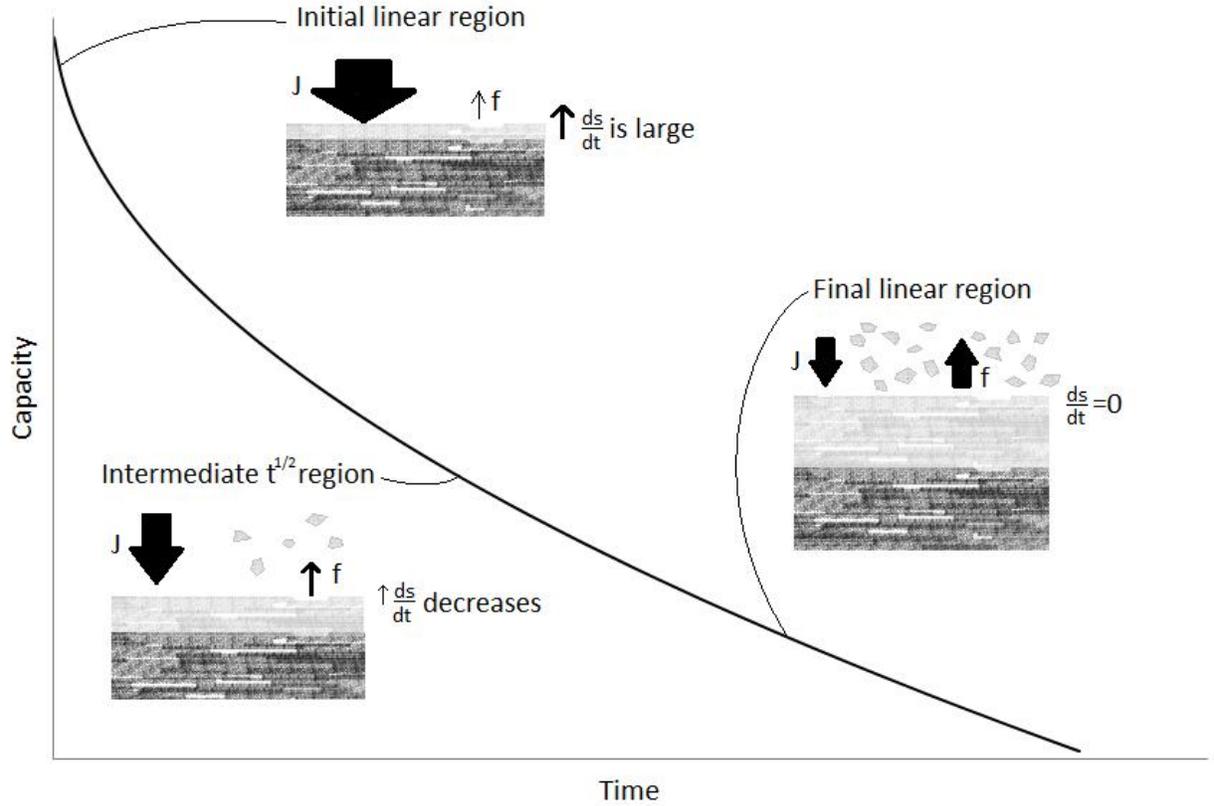

**Figure 11:** When SEI dissolution or delamination occurs, capacity fade occurs in three regimes. Initially the SEI thickness s is small, the SEI formation rate J is very large, limited only by the reaction rate, and the SEI loss rate f is small, so s increases rapidly while capacity decreases rapidly. For intermediate times, SEI formation is limited by diffusion through the existing SEI, but J is still greater than f, leading to an approximate $t^{1/2}$ rate of capacity fade. Eventually the SEI formation and loss rates will be equal, leading to constant s and a steady rate of capacity loss.

Although for most choices of $f(s)$ it is not possible to find an analytic formula for $Q(t)$, we can make progress in the simple case $f(s) = \frac{s}{t_0}$, in the limit $k \to \infty$ (which experiments[28] suggest is reasonable: see above). Under these conditions,

$$Q(t) = Q_0 - A\sqrt{\frac{\rho c D t_0}{m}}\left(\frac{t}{t_0} + \ln\left(1 + \sqrt{1 - e^{\frac{-2t}{t_0}}}\right)\right),$$

where $Q_0$ is the initial capacity. We see that the long-term linear fade has the same prefactor as the $t^{1/2}$ fade of the stable case, except for the dimensionally-required $1/(2t_0^{1/2})$. This long-term behavior is not influenced by the assumption $k \to \infty$.

**Conclusion**

We have developed a general theoretical framework to model capacity fade and lifetime statistics in rechargeable batteries, focusing on the mechanism of SEI formation at the anode. We have shown that a very simple single particle reaction model is sufficient to quantitatively understand SEI formation on graphite, since it provides a good fit to observed capacity fade. Moreover, computational results with a more complicated porous electrode model of capacity fade show that no significant spatial variations form at the cell level. The temperature dependence of the diffusivity of the limiting reacting species through SEI can be modeled using an Arrhenius dependence. This model can be used to model the degradation of rechargeable batteries in a variety of thermal conditions, to guide and interpret accelerated aging experiments, and to understand the statistics of battery lifetime. The model can also be extended to account for rapid SEI growth due to area changes and SEI loss, which occur, for example, in nanostructured silicon anodes.

Although we have focused on SEI formation in Li-ion batteries, some of our models and conclusions may have broader relevance to other battery technologies, not involving ion intercalation. It is important to emphasize that, although SEI growth leads to capacity fade, it also plays a crucial role in rechargeable battery engineering. The formation of a stable SEI layer protects the anode at large negative potentials and allows the design of high-voltage batteries operating outside the ``voltage window'' for electrochemical stability of the electrolyte. The same general principle enables the stable operation of lead-acid rechargeable batteries (> 2 V), where the half-cell reactions at the lead anode and the lead oxide cathode both form amorphous lead sulfate films, which allow active ions to pass, while suppressing water electrolysis outside the voltage window of the sulfuric acid electrolyte (~ 1 V).[48] In that case, capacity fade can also occur by irreversible reactions involving active species via lead sulfate crystallization (``sulfation'').[49–52] Although this capacity fade mechanism differs from SEI formation in Li-ion batteries in the microscopic details, the engineering consequences are the same (increased interfacial resistance and loss of active material), and it may be possible to apply and extend our models of capacity fade and lifetime prediction to this and other rechargeable battery systems.

**Acknowledgements**

This work was supported by Samsung via the Samsung-MIT Program for Materials Design in Energy Applications. The authors wish to thank Todd Ferguson for substantial computational assistance, and Michael Hess, Raymond Smith and Juhyun Song for useful discussions.

1. M. S. Whittingham, *Chemical Reviews*, **104**, 4271 (2004).

2. N. Balke, S. Jesse, A. N. Morozovska, E. Eliseev, D. W. Chung, Y. Kim, L. Adamczyk, R. E. García, N. Dudney and S. V Kalinin, *Nature Nanotechnology*, **5**, 749 (2010).

3. J. Newman and K. E. Thomas-Alyea, *Electrochemical Systems*, Third ed., Wiley-Interscience, Berkeley (2004).

4. M. Doyle, T. Fuller and J. Newman, *Journal of the Electrochemical Society*, **140**, 1526 (1993).

5. V. Srinivasan and J. Newman, *Journal of The Electrochemical Society*, **151**, A1517 (2004).

6. S. Dargaville and T. W. Farrell, *Journal of The Electrochemical Society*, **157**, A830 (2010).

7. G. K. Singh, G. Ceder and M. Z. Bazant, *Electrochimica Acta*, **53**, 7599 (2008).

8. P. Bai, D. A. Cogswell and M. Z. Bazant, *Nano Letters*, **11**, 4890 (2011).

9. D. A. Cogswell and M. Z. Bazant, *ACS Nano*, **6**, 2215 (2012).

10. M. Tang, W. C. Carter and Y.-M. Chiang, *Annual Review of Materials Research*, **40**, 501 (2010).

11. Y. V. Mikhaylik and J. R. Akridge, *Journal of The Electrochemical Society*, **151**, A1969 (2004).

12. J. Shim, K. A. Striebel and E. J. Cairns, *Journal of The Electrochemical Society*, **149**, A1321 (2002).

13. U. Kasavajjula, C. Wang and A. Appleby, *Journal of Power Sources*, **163**, 1003 (2007).

14. C. S. Wang, *Journal of The Electrochemical Society*, **145**, 2751 (1998).

15. E. V. Thomas, I. Bloom, J. P. Christophersen and V. S. Battaglia, *Journal of Power Sources*, **184**, 312 (2008).

16. H. J. Ploehn, P. Ramadass and R. E. White, *Journal of The Electrochemical Society*, **151**, A456 (2004).

17. A. J. Smith, J. C. Burns, X. Zhao, D. Xiong and J. R. Dahn, *Journal of The Electrochemical Society*, **158**, A447 (2011).

18. P. Ramadass, B. Haran, P. M. Gomadam, R. White and B. N. Popov, *Journal of The Electrochemical Society*, **151**, A196 (2004).
27